# Single-Cell Proteomics using Mass Spectrometry

Amanda Momenzadeh, Jesse G. Meyer*

Department of Computational Biomedicine, Smidt Heart Institute, Board of Governors Innovation Center, Advanced Clinical Biosystems Research Institute Cedars Sinai Medical Center, Los Angeles CA 90048

* Correspondence to Jesse.Meyer@cshs.org

ABSTRACT

Single-cell proteomics (SCP) is transforming our understanding of biological complexity by shifting from bulk proteomics—where signals are averaged over thousands of cells—to the proteome analysis of individual cells. This granular perspective reveals distinct cell states, population heterogeneity, and the underpinnings of disease pathogenesis that bulk approaches may obscure. However, SCP demands exceptional sensitivity, precise cell handling, and robust data processing to overcome the inherent challenges of analyzing picogram-level protein samples without amplification. Recent innovations in sample preparation, separations, data acquisition strategies, and specialized mass spectrometry instrumentation have substantially improved proteome coverage and throughput. Approaches that integrate complementary omics, streamline multi-step sample processing, and automate workflows through microfluidics and specialized platforms promise to further push SCP boundaries. Advances in computational methods—especially for data normalization and imputation—address the pervasive issue of missing values, enabling more reliable downstream biological interpretations. Despite these strides, higher throughput, reproducibility, and consensus best practices remain pressing needs in the field. This mini review summarizes the latest progress in SCP technology and software solutions, highlighting how closer integration of analytical, computational, and experimental strategies will facilitate deeper and broader coverage of single-cell proteomes.

INTRODUCTION

Bulk proteomics has been important for understanding diseases, but it simplifies our view of biological systems because it averages single cell protein quantities across thousands of cells. Single-cell proteomics (SCP) enables a view of heterogeneous cell states and functions, for example the cell proportions in tissues.[1,2] However, SCP is substantially more challenging due to the minuscule protein quantities per cell, vast range of protein abundance, and the fact that proteins cannot be amplified as in transcriptomics.[1] This creates challenges in protein identification and quantification, and computational workflows may not be optimized for the low-signal, sparse nature of SCP data.[2] The need for highly sensitive detection and reproducibility in SCP is essential. In SCP, bottom-up mass spectrometry is preferred for its significantly greater sensitivity and broader proteome coverage than top-down.[1] Several prior reviews have covered SCP more broadly.[1,3–9] Here we give a brief overview of recent advances in bottom-up SCP sample preparation, data acquisition and computational tools, focused on the last 2-3 years.

A general workflow for SCP is shown in **Figure 1**. A population of dissociated cells is obtained from the biological system of interest. Those cells are sorted into single cells using various sorters and various substrates. Cells are lysed mechanically or chemically before proteolysis with trypsin to generate peptides. Optionally, peptides are covalently tagged with amine-reactive multiplexing reagents that enable pooling of multiple cells. The single cells or pools of cells are then analyzed by liquid chromatography hyphenated to mass spectrometry to detect and quantify peptides, which are used to infer protein quantities.

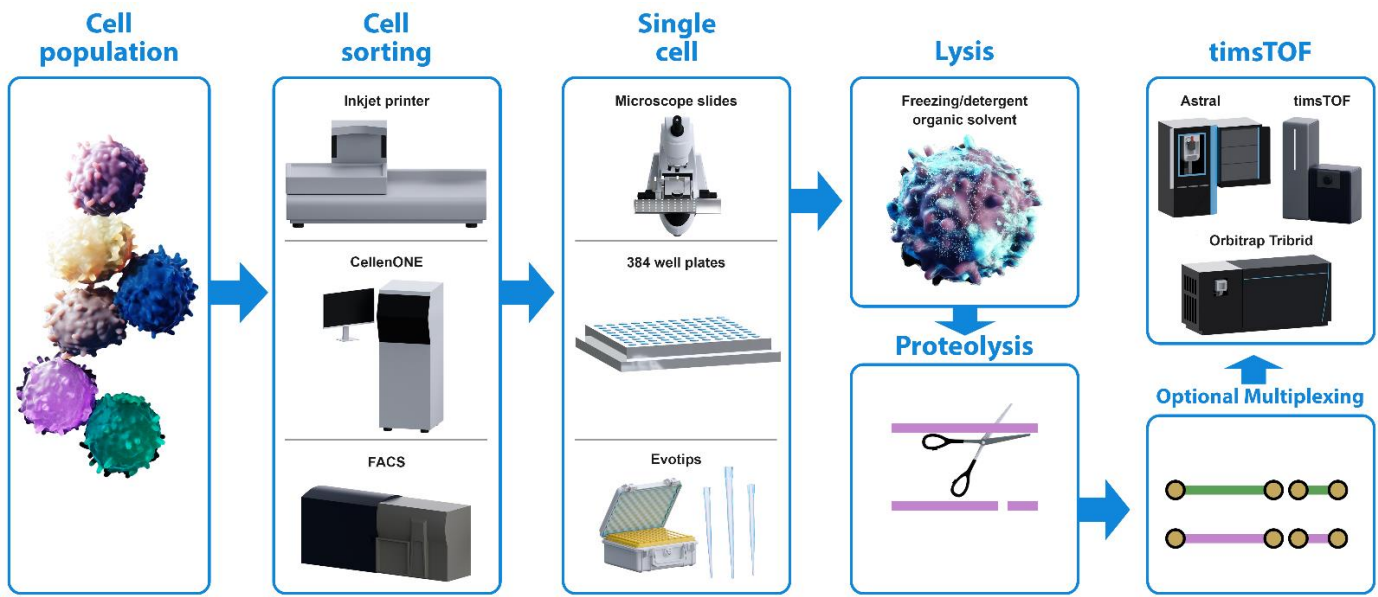

**Figure 1. General workflow for SCP-MS highlighting the choices for cell sorting, cell substrates, multiplexing, and data collection.**

Relevant to many advances discussed in this review, there are two competing data collection strategies for single cell proteomics: Data-Independent Acquisition (DIA) and Tandem Mass Tag (TMT) multiplexing. These two distinct mass spectrometry approaches differ primarily in their data acquisition and quantification strategies. TMT relies on isobaric chemical labeling, where peptides from multiple samples (i.e., individual single cells) are tagged with mass-encoded reporter ions and pooled for simultaneous analysis in a single mass spectrometry run[10] (**Figure 2A**). Quantification in TMT is achieved by measuring the relative intensities of these reporter ions in MS2 or MS3 scans, allowing for multiplexed sample processing with up to 35 channels in modern workflows. In contrast, DIA is a label-free method that systematically fragments all precursor ions within a specified mass range, capturing comprehensive MS/MS spectra in a single run[11] (**Figure 2B**). This approach enables continuous and unbiased peptide sampling, reducing the stochastic nature of Data-Dependent Acquisition (DDA) and enhancing quantitative reproducibility. While DIA directly measures peptide abundances within each sample, TMT relies on comparative quantification across multiplexed samples, requiring additional computational corrections to address signal interference.

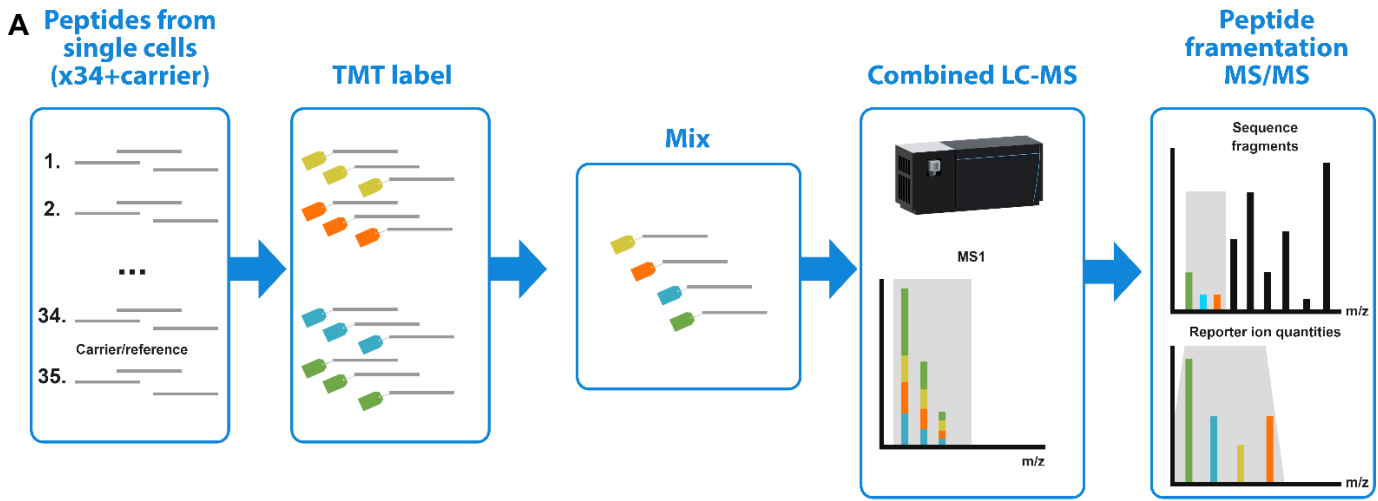

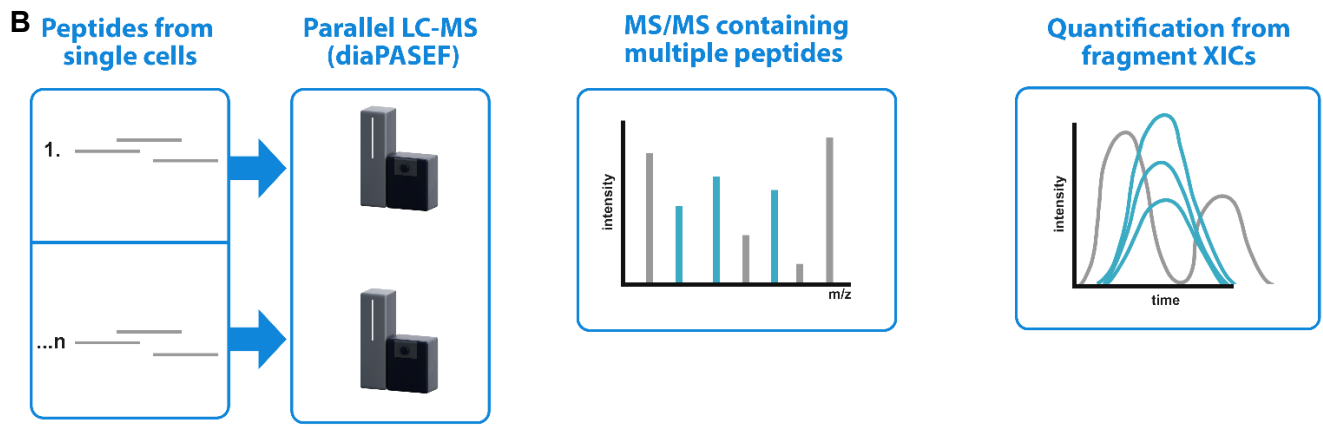

**Figure 2: Comparing the two main competing SCP-MS strategies with TMT and DIA.**

DIA and TMT each have strengths and weaknesses when applied to single-cell proteomics. TMT excels in throughput by allowing multiple single cells to be analyzed in parallel within a single LC-MS run, reducing instrument time per sample.. Additionally, the use of a carrier proteome, where an excess amount of pooled protein is included in one TMT channel, enhances peptide identification for low-abundance proteins. However, TMT suffers from ratio compression and co-isolation interference, where peptides from different samples share precursor ions and fragment together, distorting quantification and reducing dynamic range. Furthermore, ion suppression effects from highly multiplexed samples, especially in the presence of a carrier, can hinder the detection of low-abundance proteins, limiting sensitivity. TMT may also experience issues with missing values across conditions due to the stochasticity of peptide fragmentation across TMT batches; this problem that is largely alleviated with prioritized SCoPE (pSCoPE), where peptides of interest are preferred for fragmetnation across batches[12]. DIA-LFQ, on the other hand, avoids these issues by independently measuring each sample's peptide abundances, eliminating inter-sample interference. This leads to more accurate, complete and reproducible quantification, a wider dynamic range, and improved sensitivity for low-copy proteins. However, DIA requires a separate LC-MS run per cell or a small pool of cells, making its throughput in terms of cells per unit of time lower than TMT. Nevertheless, recent advances in DIA instrumentation, MS1-based multiplexing[13,14], and data analysis pipelines have significantly increased the scalability of DIA, making it a powerful alternative for unbiased single-cell proteomics. Ultimately, the choice between DIA and TMT depends on the trade-off between throughput and quantitative accuracy, with TMT excelling in high-throughput settings where multiplexing is crucial, while DIA-LFQ is increasingly favored for its superior quantification, sensitivity, and dynamic range.

A broad survey of SCP figures of merit reveals wide variation in performance across recent studies. We tabulated the number of proteins quantified per minute, proteins quantified per cell, corrected run time in minutes/cell (lower is better), and the data acquisition type (**Figure 3, Supplemental Table 1**). We found that TMT methods consistently quantify the most proteins per minute owing to their multiplexing of multiple cells per injection, with the 32-plex TMT leading. Accordingly, TMT also produces the lowest analysis time per cell. The most proteins per cell (x-axis) were found in studies using the Astral, where over 5,000 proteins could be detected. Most studies used DIA methods. Future studies should focus on filling in the upper right quadrant with high proteins per cell and high proteins quantified per minute. In agreement with a recent perspective, we propose that a main limitation requiring further advances is the need for higher throughput to better sample cell populations.[15]

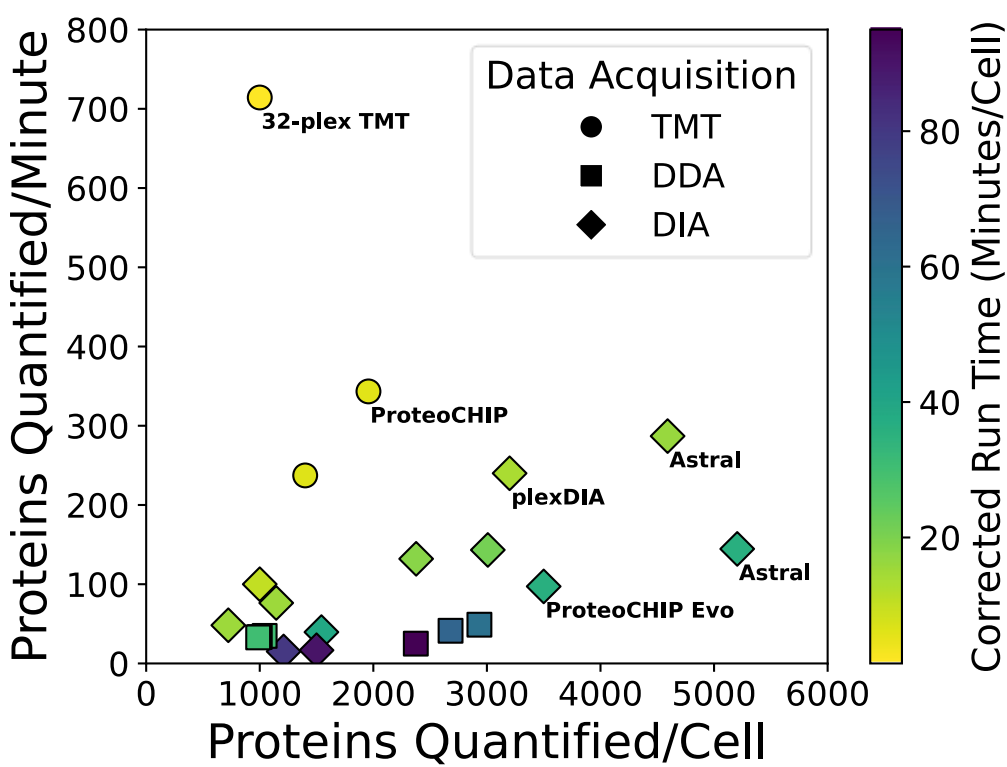

**Figure 3: A random survey of SCP figures of merit across studies from the last two years.** Proteins quantified per minute (y-axis) is plotted against Proteins quantified per cell (x-axis). The corrected run time in minutes of data collection per cell is shown as the color mapping, and acquisition type is given as the marker shape.

**Sample preparation**

Experimental design in single cell studies is critical to avoid technical and batch artifacts, such that biological effects can be distinguished with confidence. Incorporating a consistent reference sample across batches helps to control for signal shifts, especially if the reference closely resembles the study samples and includes all cell types of interest.[8,16]

For inter-group or longitudinal comparisons, distributing samples randomly across batches uniformly can further minimize batch-related variability. If this randomization was not done, one should include batch as a variable in models.

A typical proteomics experiment aims to inject one microgram of total peptide into the LC-MS system. The average single mammalian cells contain 5000-fold less protein, about 200 pg.[4] Adapting sample preparation approaches that were designed for microgram scale quantities to the ~200 pg is a substantial challenge that continues to be a major research focus in the SCP field. For example, at these ultra-low sample amounts, surface absorption to pipette tips and tube walls can cause noticeable losses in observable proteomic depth.

Additional challenges include the need to sort cells. While many studies have successfully used a traditional fluorescence activated cell sorter, more recently, the CellenONE instrument has become commonplace as it images each single cell and offers flexible dispensing patterns. An innovation solution is the recent adaptation of an inkjet printer for single cell dispensing and reagent dispensing in 384 well plates.[17,18] This approach can offer considerable cost savings while still providing fast and reproducible sorting, achieving up to 2,600 protein identifications with label free quantification.

The chemical choices for lysis and proteolysis solutions also require unique considerations for SCP. An extensive exploration of one-pot sample preparation and data collection was recently performed by Matzinger *et al*.[19] This work is an important read for new SCP practitioners because it covers nearly all aspects of sample preparation and it presents a simple protocol for single cell sample preparation in 384 well plates. Among the many conclusions, they report optimal isolation parameters for CellenONE, choice of trypsin, and various data analysis options. Other best practices include minimizing sample transfers (even a single transfer can reduce protein identification by 50%), keeping samples hydrated with DMSO during prior to placing in the autosampler, adding trypsin in two boluses rather than a single addition, and use of short 5.5 cm µPAC columns rather than packed beds.

Given the importance of limiting surface contact during SCP sample preparation, there have been several studies on the topic of novel surfaces. A recent concept commercialized by CellenONE is proteoCHIP,[20] which is a collection of nanowells covered with oil. After cell dispensing, nanoliter lysis, digestion, and TMT labeling, the cells in one group are pooled with a funnel device by centrifugation and then directly placed into the autosampler. This enabled identification of nearly 2,000 protein groups per TMT plex injection using a 20x carrier. This approach substantially reduces the amount of manual manipulation required for single cell TMT experiments at the added expense of additional consumables.

A subsequent study described proteoCHIP EVO. This device arranges single cells on pedestals such that they can be directly centrifuged into a 96 tip rack of EvoSep One tips for analysis.[21] The authors found that this direct loading approach resulted in about twice as many protein identification compared to manually transferring cells to an autosampler vial. However, these results may be confounded by losses solely during the sample transfer process, which is known to cause loss of nearly half of proteins.[19] While a direct comparison with one-pot 384 well sample preparation is needed, this approach represents an easy and reproducible advance to automate the introduction of cells into a mass spectrometer. Use of EvoTips may present a return on cost investment, as these protect the analytical column and limit accumulation of contaminants on trap columns, which may otherwise degrade over thousands of single cell injections.

Another new sample preparation idea is called nano-proteomic sample preparation (nPOP),[22] which uses fluorocarbon-coated microscope glass slides as a flexible surface for any sample layout. Nanoliters of DMSO are deposited onto the slide followed by a single cell, then digestion mixture, and optionally, a multiplexing label. A larger droplet is then used to collect any cell droplets of interest and move them to an autosampler-compatible 384 well plate. This approach was used to explore functional protein covariation across single cells related to the cell cycle.[23]

In line with limiting sample contact with surfaces, a recent report described using a microfluidic liquid handling robot for cell sorting and sample preparation, directly in autosampler vial inserts.[24] The authors achieved over 3,000 protein groups on average in HeLa and A549 cells using DIA and match between runs (MBR). A benefit of this approach is that the cell sampling robot can sample specific single cells of interest from a culture plate; for example, the authors demonstrated single cell sampling of cells that start migration into the cleared zone in a plate scratch assay.

Automating sample preparation processes with microfluidics is an active area of research. Microfluidics is a versatile, multidisciplinary technology that enables precise manipulation, observation, and processing of fluid samples ranging from nanoliters to femtoliters into well-defined compartments.[25] As proteomic analysis shifts towards the single-cell level – where precise and reproducible sample preparation is paramount – techniques for handling low-input samples and liquid volumes are becoming increasingly critical. Innovative methods such as acoustic droplet ejection and digital microfluidics allow for precise separation and manipulation of small volumes while conserving reagents and minimizing waste. These approaches are particularly well-suited to enhancing the sensitivity and accuracy of single-cell analyses. For instance, Peng and colleagues introduced an all-in-one digital microfluidic pipeline that integrates proteomic sample reduction, alkylation, digestion, isotopic labeling, and analysis.[26] A recent advance demonstrated that cell capture, lysis, digestion, and peptide separation can be carried out in an integrated microfluidic chip.[27] This approach was able to quantify about 1500 proteins from each single cell in a 20 parallel-cell chip. This depth is generally aligned with that expected using similar separation settings with an Orbitrap.

Integrated platforms like CellenONE,[22,28–30] which combine cell sorting and small-volume liquid dispensing, have also been utilized for cell separation and sample preparation, further advancing single-cell proteomics automation and standardization. However, in scenarios where the cell sorting equipment is at a remote institution or if samples are collected asynchronously, it, may be difficult to sort cells fresh for every experiment. To determine the effects of various storage conditions, Onat *et al.* recently investigated how cell refrigeration or freezing would influence the observed SCP profiles.[28] The authors found that either storage method induced cell size and proteome deviations from the freshly prepared control cells. Notable changes were observed in oxidative phosphorylation and translation pathways, among several other pathways. They concluded that to capture endogenous biology, fresh cells are required, but still much information is obtainable from stored cells.

Proteomic sample proteolysis often requires multiple steps, such as additions of lysis and solubilization agents, followed by reducing agents, alkylation of free cysteines, and occasionally separate proteolysis with LysC and trypsin. A recent study showed that a single addition of a one-step cocktail (*i.e.,* omitting sequential lysis, reduction, alkylation, and proteolysis) could achieve results that were identical to sequential additions.[31] They also showed that addition of a reducing agent greatly reduced peptide identifications, possibly due to denaturation of the proteases by reduction of their internal disulfide bonds. Overall, the trend in the SCP field is to use simple one-step sample processing to obtain peptides for mass spectrometry.

**Innovations in data collection**

One of the early strategies for single cell proteomics leveraged the signal stacking of TMT plus a carrier channel to achieve the necessary sensitivity, even with old hardware. This MS2-based quantification has limitations for the large-scale projects required for SCP where thousands of cells must be analyzed. In many ways, DIA is simpler and provides more complete data across samples. However, although multiplexing with DIA has been reported for relative quantification of peptides against heavy signals, it is not widely used.[32,33] Recently, two studies reported ideas to resurrect 15 year old MS1 based multiplexing techniques to pair them with DIA for SCP multiplexing. Both approaches allow using one channel as a carrier to boost sensitivity by limiting absorption losses.

One approach, plexDIA,[14] uses three plex peptide labeling with the original mTRAQ reagent[34] to achieve triple the throughput for low sample amounts. They found that with 500 ng sample loads and the first-generation Q-Exactive Orbitrap, they could quantify about 6,400 proteins per injection in 60 min separations (excluding sample loading and column regeneration). The quantitative accuracy was similar to that of separate DIA injections. From single cells, they were able to quantify about 1,000 protein groups from each of the three plexes using a timsTOF SCP and a 30 min total runtime. The data completeness using this approach is high at 98% across plexes. This data collection advance was paired with updates to the DIA-NN software to leverage this new type of data. This advance may enable DIA multiplexing to larger scale SCP projects.

The second approach uses a different label that was also common fifteen years ago called dimethyl labeling.[34] Depending on the combinations of deuterium and carbon used for dimethyl labeling, normally 3-plex can be achieved. Their approach pairing dimethyl labeling with DIA was called mDIA.[13] They observed a slight reduction in total protein identifications and quantitative precision at the bulk proteome level going from unlabeled DIA to three plex. They also demonstrate that by using Lys-N they can achieve two N-terminal dimethyl events per peptide which increase the plex from three to five. To compute relative quantities from this data, they developed an approach called RefQuant that samples the mean of the 40% lowest quantile of all available ion ratios. Finally, they showed that using this approach with one reference channel and two single cells doubled the number of protein groups they can identify in single cells relative to results from their prior publication to a median of 2,377 using a timsTOF SCP with EvoSep One Whisper 40 samples per day.

A challenge with the common Single Cell ProtEomics by Mass Spectrometry (SCoPE-MS) approach using TMT multiplexing is that it benefits from MS/MS measurement of the same peptides across TMT batches, which is always less than 100% due to the stochasticity of DDA. Prioritized mass spectrometry overcomes this problem by preferentially fragmenting certain precursor *m/z* species that are known to be of interest *a priori*.[12] They achieved this with the MaxQuant.Live software to operate the mass spectrometer. Once the high priority list is exhausted, lower priority peptides are then targeted. This approach increased data completeness for the 4,000 high priority peptides to 72%, compared to 49% without prioritization, while also doubling the number of proteins identified per cell. The utility of this approach was demonstrated by applying it to study macrophage polarization.

A new concept in SCP is called wide window acquisition (WWA) with data dependent acquisition (DDA). This embraces the idea of peptide co-isolation in the same quadrupole window as a strength by further increasing the window width and then using software that can find multiple precursors, along with match between runs, for MS1 level quantification. This was first reported with low nanoliter flow LC.[35] Using 200 pg aliquots, they found that WWA could identify slightly more proteins than standard DDA or DIA. However, the proteins quantified from DIA exhibited higher quantitative precision relative to those with DDA or WWA. In a subsequent paper, WWA was found to increase up to 150% more proteins than narrow window isolation, but results from DIA were not reported for direct comparison.[36] In agreement with the first paper, with the single cell sample loads, there was no difference in the number of quantified proteins when considering MBR from the multiple normal DDA files *versus* WWA. Overall, while an interesting concept that warrants further exploration, based on the presented data in these two papers, the current results from WWA do not provide a clear benefit over simple DIA.

For DIA, there are avenues to optimize and adapt DIA to single cells, especially given that Orbitrap instruments offer great flexibility in possible scan sequences that achieve variable resolution and measurements per chromatographic peak. A study by Petrosius *et al.* found many such optimizations.[37] First, in contrast to high load DIA where smaller windows convey selectivity, wider DIA isolation windows improved sensitivity for low input. For 100 ng inputs, optimal identifications were obtained with 20 *m/z* windows, while for 1 ng inputs, 80 *m/z* windows were best. They also found that using high resolution MS1 scans at regular spacing before completing the qualitative MS2 scan sequence improved identification and quantitative precision. When coupled with µPAC columns, this workflow resulted in over 1,461 proteins in 20 min. Adding a high load library for identification transfer appeared beneficial, and they found that this method could reveal heterogeneity in single stem cells.

Another DIA idea focused on only MS1 data collection inspired by the accurate mass and time approach is called transferring identifications based on FAIMS filtering (TIFF).[38] The approach first generates a spectral library using FAIMS fractions, which was found to increase the ion injection times and signal-to-noise ratio. The 3D matching based on retention time, MS1 *m/z*, and FAIMS mobility then enabled peptide identification. This identification approach was assessed using an entrapment approach with a bacterial library. The 3D matching approach reduced the bacterial peptide false matching rate to 1.8%, as compared with 4.1% when using 2D matching. They achieved 1,210 protein identifications from HeLa using 80 min total injection times.

**Separations**

Trap columns are often used with SCP because they can speed up sample introduction and help with the lack of offline cleanup used for single cells. However, trap columns are also prone to clogging on the scale of thousands of single cell samples. A new idea is to leverage tubular open trap columns that bind their analytes on the walls due to diffusion, thus allowing particulate and larger aggregates to flow straight through.[39] This idea further simplifies sample introduction because it can serve as a sample loop in addition to reducing the back pressure substantially.

While trap columns do reduce the overhead between data collection by allowing high flow transfer of samples to analytical columns, further modifications offer additional time savings. A recent study reported eliminating run-to-run overhead to maximize high throughput SCP by using a dual trap configuration. In this dual trap single column (DTSC)[40] approach, one trap is loaded and washed while the second is eluted to the analytical column.[29] The DTSC approach enabled profiling of hundreds of single cells and identification of hundreds of proteins from single cardiomyocytes and single cells from dissociated aortas.

An equivalent idea is to have multiple columns.[41] In this approach, one column performs the online separation to the mass spectrometer while another column is being regenerated for the next injection. The authors found that this approach can reduce total analysis times down to 7 min per cell, allowing over 200 cells per day and an average of 621 protein group identifications. To eliminate the added cost of a second binary pump normally required for this method, several variations such as offline gradient generation have been reported.[42]

**Instrumentation**

The first commercial breakthrough to greatly increase MS sensitivity and enable commercial scale SCP was the timsTOF. The timsTOF leverages TIMS to preconcentrate peptide ions as they elute from the column while separating the previous batch of concentrated ions, increasing signal by about four-fold[43] and enabled detection of 1,000 to 2,000 proteins from single cells. Since then, two updated instruments with higher sensitivity have been release (timsTOF Ultra and Ultra 2). The latter claims the ability to measure over 1,000 proteins from just 25 pg.

Another major advance of instrumentation that will enable SCP is the Astral mass analyzer.[44] This sensitive and fast mass analyzer enables extremely specific peptide analysis with small quadrupole isolation windows. It was found to double the number of peptides compared to the Orbitrap Eclipse with 1 ng sample loading (~31,000 peptides with Astral *versus* ~16,000 with Eclipse). After further scan sequence optimization, ~4,000 proteins could be identified from 250 pg aliquots of HeLa, and ~3,500 proteins could be identified from single HEK293 cells with 22.5 min total runtime in a 384-well plate format. A subsequent paper suggested that combining the Astral with the ProteoCHIP mentioned earlier can boost single cell protein group identifications to over 5,000, which would be a landmark in single cell depth if widely reproducible.[45]

**Spectral Library Considerations for SCP**

DIA-based SCP has become common since the first reports of SCoPE based on TMT. There are several challenges present with SCP peptide spectrum matching (PSM) that are not present when dealing with bulk proteomics samples. The quality and comprehensiveness of the library directly influence the accuracy and number of PSMs. Some of these issues relate to false discovery rate calculation, where the number of decoys influences the outcome. Perhaps counterintuitively, if a library is too large, then it will increase decoy matches and drive down total identification. Whereas if the library contains a smaller set of proteins that better match the true proteins obtainable from that sample, then more proteins can be identified. Another spectral library consideration is that using DDA peptide identification, SCP DDA spectra from Orbitraps have fewer annotated fragment ion peaks and a lower signal-to-noise ratio.[2] This problem does not appear to be true for timsTOF data.[46] This suggests that, at least for Orbitrap DIA SCP methods, building libraries from other SCP data may be beneficial.

There have been many suggestions about tailoring the spectral library used for SCP. A recent paper described how adding "matching enhancer" injections with slightly higher input material and using the "match between runs" feature of DIA-NN can improve SCP depth by 16% per individual cell.[47] They extensively explored the false transfer rate using *E. coli* spike in and found the metrics to be suitable. They demonstrated how this could reveal single cell heterogeneity in U-2 OS cells treated with interferon gamma.

**SCP Data Processing and Interpretation Resources**

The explosion of SCP has led to some new data processing tools that are specific for SCP data. **Table 1** summarizes those tools.

SCP enables the identification of specific cell populations and provides an understanding into their differentiation trajectories.[48] After protein quantities are obtained from SCP data, the first step is often to summarize the cell types and their markers using established workflows from scRNA-seq such as scanpy[49] in Python or Seurat[50] in R. However, there is no standardized, widely accepted pipeline for SCP data pre-processing and studies use different analysis methods in varying orders. A unique challenge in SC is that cells vary in size, leading to substantial differences in total protein content across individual cells. Normalization methods applied in bulk proteomics, such as subtracting the median from log-transformed protein quantities, are unsuitable when there are vast differences in total protein content across single cells.[8] Instead, Gatto *et al.* recommend normalization based on a metric that reflects cell size, such as total protein content or a common reference protein.[48] The order of processing steps must also be carefully considered to ensure the assumptions of each step are met. For example, a *t*-test requires prior batch correction, whereas linear regression can include technical factors like batch as a variable in the model, eliminating the need for prior batch correction.[48]

SCP-specific R packages have been introduced in the past two years. One called scp,[51] offers a comprehensive pipeline that includes key steps such as quality control (at both feature and cell levels), data aggregation (PSMs to peptides to proteins), normalization, and batch correction. The scp package supports data input from MaxQuant, Proteome Discoverer, DIA-NN, and requires a sample annotation table for processing. Implementing two different data processing workflows, SCoPE2 and SCeptre, using scp revealed workflow dependent differences in clustering, which could lead to varying biological interpretations.[48] Another package, scplainer[52] builds on scp capabilities by integrating SCP data processing steps with linear modeling to facilitate deeper biological insights. There is a need for a community effort to replicate benchmarking experiments across various SCP protocols to better assess the robustness of computational workflows.

Additionally, these tools require some knowledge of programming, which is a barrier for many mass spectrometrists and biologists. The Meyer lab recently published a new platform that enables no-code interpretive single cell proteomics data analysis called the platform for single cell science (PSCS).[53] This platform is unique in that it enables reproducible analysis through containerization, which is hidden from the user interface. It also allows collaboration where multiple users can be added to a project to run additional analyses. Importantly, it allows interactive data exploration using the CellXGene JavaScript interface in the browser. Finally, it has a unique publication mechanism where users can share their input data of cells and quantified genes, their exact analysis pipeline, the results that were generated including figures and CellXGene interactive exploration, and the final annotated data object for those that may want to explore further. Importantly, the publication page displays the exact analysis pipeline for inspection in the no-code interface, and then anyone can clone that pipeline for their own project, which maximizes the reproducibility and transparency of SCP data analysis.

Biological (*e.g.* vast differences in the number of proteins per cell due to cell size differences) and technical (*e.g.* presence of batch effects due to typically large scale of single cell MS acquisitions[54]) factors result in a significantly higher prevalence of missing values in SCP compared to bulk proteomics. In fact, 50–90% missing values per single cell is common, whereas bulk proteomics typically has a maximum of 50% missing values.[54] Imputation methods designed for scRNA-Seq may not be suitable for SCP because missing values differ fundamentally between the two: zeros in scRNA-Seq are biologically meaningful, while zeros in SCP often result from detection or computational limitations, necessitating tailored imputation strategies for SCP.[54]

To handle missing data effectively, imputation methods for low coverage MS data using LFQ have emerged over the last two years. PIRAT (Precursor or Peptide Imputation under Random Truncation), for example, addresses challenges in processing protein groups for which only a single peptide is identified.[55] By leveraging correlations among peptides from the same protein, it handles both random and censored missing values within a single statistical model. Similarly, PIMMS (Proteomics Imputation Modeling Mass Spectrometry) employs three deep learning models–collaborative filtering, a denoising autoencoder and variational autoencoder–to impute missing values.[56] These three models performed comparably on simulated data, recovering signals across the entire distribution including low abundance features, and scaled effectively to high-dimensional datasets. Additionally, unlike heuristic methods, these methods produced conservative imputations without biasing values toward detection limits.

**Table 1. Single cell proteomics data analysis tools**

| Tool name | Purpose |
|---|---|
| Scanpy[49] | All-purpose single cell python package |
| Seurat[57] | All-purpose single cell R package |
| PSCS[53] | No-code interface to multiple tools |
| scp[51] | SCP-specific methods such as precursor rollup to proteins |
| scplainer[52] | variance analysis, differential abundance analysis and component analysis |
| PIRAT[55] | Statistical imputation for proteomics |
| PIMMS[56] | Imputation with deep learning |
| SCeptre[58] | Extends scanpy for SCP data with normalization of TMT data and outlier removal |

**Single cell Multi-Omics Integration**

There is much interest in pairing single cell proteomics with other single cell -omics, especially single cell genomics, which leads to several new analytical challenges. Fulcher *et al.* show that surface tension can partition single cell lysis droplets in half, such that two separate plates can be generated for either SCP or scRNA-seq.[59] This approach, appropriately named nanoSPLITs, allows for quantifying an average of about 6,000 transcripts and 3,000 proteins from a single cell. A longstanding question surrounding proteome/transcriptome integration is whether poor correlation from bulk samples is due to the averaging of many cell types. Here, they found that from single cells, the correlation is still poor, suggesting that variation is likely due to differing rates of transcription, translation and protein degradation. They also found there is generally poor agreement between the protein or transcript marker. Finally, they showed that they can apply this technique to cells dissociated from human pancreatic islets and how they could use label transfer from reference maps to determine cell types. This allows comparison of protein quantities from single cells according to defined RNA markers.

The study that introduced the timsTOF SCP mass spectrometer provided key insights between differences in single cell proteomics and single cell transcriptomics[43]. Notably, like Fulcher et al., the research found a low correlation between single-cell proteomes and transcriptomes, indicating that protein levels cannot be reliably predicted from RNA levels even in individual cells. While transcript expression showed good correlation between different scRNA-seq technologies, protein measurements diverged significantly from RNA data in PCA, and cell-cell correlations were higher on average within the proteome compared to transcriptomes. Furthermore, the study observed differences in data completeness, with single-cell proteomics achieving higher gene or protein expression completeness than both SMART-Seq2 and Drop-seq. Analysis of variability revealed that the single-cell transcriptome exhibits a much higher overall stochasticity compared to the proteome, likely due to shot noise limitations at the RNA level, while protein variability appeared more linked to measurement sensitivity. Interestingly, the researchers identified a stable core proteome comprising the 200 least proteins, whose corresponding transcripts showed a wide range of variability, suggesting substantial post-transcriptional regulation. This finding underscores that proteomics provides complementary information to transcriptomics, as protein and RNA levels are regulated differently in single cells. The study emphasizes that direct protein measurements are crucial for a comprehensive understanding of cellular states and the complex interplay between mRNA and protein abundance. Ultimately, this work highlights the unique insights gained from high-sensitivity single-cell proteomics, revealing aspects of cellular heterogeneity and regulation that are not evident from transcriptomic data alone and emphasizing the importance of integrating these approaches for a more complete biological understanding.

The 2021 paper that focused on the development of the SCoPE2 method contained a major substory about the study of heterogeneity in macrophages differentiated from monocytes using the U937 cell line[60]. The researchers also performed parallel single-cell RNA sequencing (scRNA-seq) using 10× Genomics to enable an integrated analysis of protein and transcript levels. Notably, as illustrated in Figure 6a, SCoPE2 was able to sample 10–100-fold more copies per gene at the protein level compared to the number of unique barcodes reads per mRNA obtained by 10× Genomics. This suggests that proteins are often present at much higher copy numbers per cell than their corresponding transcripts, allowing for more reliable counting statistics in proteomics measurements. Consistent with the Fulcher et al. and Brunner et al. papers, this study also found that RNA and protein levels in single cells often exhibit distinct patterns, suggesting widespread post-transcriptional regulation. By comparing the correlation structures of RNA and protein data, the study identified groups of genes with either similar or opposite covariation, revealing different underlying regulatory mechanisms. Genes in cluster 1 generally showed similar abundance profiles at both the RNA and protein levels and were enriched for functions including antigen presentation, cell adhesion, cell proliferation, and protein synthesis. In contrast, genes in cluster 2 displayed opposite RNA and protein profiles and were enriched for functions such as Rab GTPase activity and protein complex assembly. Furthermore, the joint analysis of cells using Conos demonstrated that while both transcriptomic and proteomic data could distinguish cell types (monocytes and macrophages), the proteomic data tended to form more compact clusters, potentially indicating lower overall variability in protein expression compared to RNA in this context. Investigating individual genes, such as p53, revealed instances where protein levels better reflected expected regulatory activity than mRNA levels, further emphasizing the importance of direct protein measurements for understanding cellular function. Overall, the findings from this study reinforce the idea that single-cell proteomics and transcriptomics provide complementary and crucial information for dissecting cellular heterogeneity and the intricate layers of gene regulation.

Hutton et al. (2024) leveraged the SCoPE2 dataset within their newly developed Platform for Single-Cell Science (PSCS) to showcase the platform's utility in facilitating reproducible and extensible analyses of single-cell multi-omics data[53]. Within PSCS, the researchers first imported the SCoPE2 data and validated the platform by recreating the original finding that transcriptomic and proteomic data could distinguish monocytes from macrophages. They then proceeded to construct a more advanced, custom analysis pipeline using PSCS's no-code interface, integrating modules from both Scanpy and CAPITAL[61]. A key aspect of their analysis was the application of the CAPITAL package within the PSCS framework to perform trajectory inference and pseudotime alignment on the combined proteomic and transcriptomic data from the SCoPE2 study. This enabled them to investigate the dynamic changes in protein and RNA levels as cells transition towards a macrophage state, identifying specific transcripts and proteins that exhibited either coordinated or disparate changes along the inferred differentiation trajectory, such as the concordant increase of CAPG at both RNA and protein levels. By demonstrating these capabilities and making the entire analytical pipeline, data, and results publicly available on PSCS (https://pscs.xods.org/p/SzFKQ), the authors highlighted the platform's potential to enhance the reusability of complex analyses and foster collaboration within the single-cell omics research community, extending the insights gained from the original SCoPE2 study through advanced temporal analysis.

The integration of single-cell proteomics and transcriptomics recently provided a more complete molecular characterization of aortic cells in a Marfan mouse model[62]. Researchers performed single-cell proteomics and integrated their data with a previously published single-cell RNA-seq dataset from a similar experiment. Initial comparisons using Seurat anchor marker analysis revealed concordance in the identification of major cell types but a lower degree of overlap when examining more refined cellular subtypes, highlighting the potential for each omics layer to capture distinct aspects of cellular identity. To further explore the relationship between the proteome and transcriptome at the single-cell level, the study employed LIGER for multi-omic integration, which allowed for the combined analysis of both data types. Despite this integration, the results indicated that the two datasets were primarily aligned at the level of general cell types, with more subtle cell sub-phenotypes not showing clear correspondence across the integrated clusters. However, a more focused multi-omic analysis specifically on smooth muscle cells (SMCs) identified unique clusters defined by both RNA and protein features, revealing Marfan-specific shifts in cell proportions within these integrated SMC clusters. Notably, multi-omic clusters enriched in wild-type cells showed higher expression of genes related to the contractile apparatus, while clusters with a greater contribution of Marfan cells exhibited enrichment in pathways related to angiotensin, TGFβ, and protease signaling, with specific markers like Lrp1 and Ace being differentially expressed. Furthermore, the presence of Tpm4, a marker of SMC dedifferentiation, in Marfan-enriched clusters alongside Ace suggests a coordinated change at both the RNA and protein levels for this phenotype. These findings underscore the complexity of integrating single-cell proteomics and transcriptomics data, where while major cell identities may be conserved, the nuanced information captured by each modality can reveal distinct and complementary insights into cellular states and disease-related alterations. The identification of specific multi-omic signatures associated with the Marfan phenotype highlights the potential of such integrated analyses to uncover novel pathways and regulatory mechanisms that may not be apparent from examining either data type in isolation.

**General recommendations for SCP**

To strengthen downstream biological interpretations of SCP MS data, if at all possible, data collection should also include all available metadata, such as the location of the cell in the tissue and phenotypic characteristics such as cell images and functional assay results.[8] Bulk cell lysates should be included as positive controls; one should count the number of cells present and then serial dilute samples to the SC level. Dispersing these bulk samples among single cell samples on the same plate, however, risks carryover of peptides. Thus, analytical columns should be verified to be clear of carryover between samples. Negative controls, prepared identically to single cell samples but without any cells, should also be included to detect potential contamination. Due to low protein quantities in SCs, the sample preparation volume should be reduced to 2-20 nL to reduce the effect of contaminants present in reagents.[8,23] Batch sizes should be as large as possible and the use of liquid handling tools, such as CellenONE[63] or nPOP[23], are recommended to reduce technical variability.

## CONCLUSIONS

Single-cell proteomics has reached an inflection point where technology, methodology, and computational advancements jointly elevate both the depth and breadth of protein profiling at the single-cell level. The continued evolution of sample handling—ranging from microfluidic devices and inkjet-based cell dispensers to new surface chemistries—has significantly curtailed sample losses. Meanwhile, DIA and multiplexing strategies expedite analysis and improve data completeness, especially when complemented by carefully curated spectral libraries. Emerging hardware innovations, exemplified by the Astral mass analyzer, promise further gains in sensitivity and throughput, propelling SCP beyond proof-of-concept studies into broader biological and clinical applications.

Still, challenges remain. High throughput and robust reproducibility are essential for capturing the full gamut of cell states and drawing reliable biological conclusions. Computational pipelines require ongoing refinement to handle the high rate of missing values and to integrate multi-omic data seamlessly. Open-source software packages and no-code platforms facilitate more transparent, reproducible pipelines, but a universal community standard has yet to emerge. By coupling state-of-the-art instrumentation with rigorous experimental design, thoughtful data analysis, and a collaborative drive to establish best practices, single-cell proteomics is poised to illuminate cellular heterogeneity at a new level of resolution and transform our understanding of biological complexity.

## AUTHOR INFORMATION

**Corresponding Author**

* jesse.meyer@cshs.org

**Author Contributions**

The manuscript was written through contributions of all authors. All authors have given approval to the final version of the manuscript.

## ACKNOWLEDGMENT

This work was supported by the National Institute of General Medical Sciences (NIGMS) R35GM142502.